\documentclass{appolb}
\usepackage{epsfig}

\makeatletter
\newsavebox{\@brx}
\newcommand{\llangle}[1][]{\savebox{\@brx}{\(\m@th{#1\langle}\)}%
  \mathopen{\copy\@brx\mkern2mu\kern-0.9\wd\@brx\usebox{\@brx}}}
\newcommand{\rrangle}[1][]{\savebox{\@brx}{\(\m@th{#1\rangle}\)}%
  \mathclose{\copy\@brx\mkern2mu\kern-0.9\wd\@brx\usebox{\@brx}}}
\makeatother

\begin{document}
\title{
\begin{picture}(0,0)(0,0)%
    \put(250,50){\makebox(0,0)[l]{\textnormal{\normalsize J-PARC-TH-0069}}}%
    \end{picture}%
Understanding experimentally-observed fluctuations%
\thanks{Presented at Critical Point and Onset of Deconfinement 2016 (CPOD2016)}%
}
\author{Masakiyo Kitazawa 
\address{Department of Physics, Osaka University, 
  Toyonaka, Osaka 560-0043, Japan}
\address{J-PARC Branch, KEK Theory Center,
  Institute of Particle and Nuclear Studies, KEK,
  203-1, Shirakata, Tokai, Ibaraki, 319-1106, Japan}
\and
Masayuki Asakawa
\address{Department of Physics, Osaka University, 
  Toyonaka, Osaka 560-0043, Japan}
}
\maketitle
\begin{abstract}
We discuss two topics on the experimental measurements of 
fluctuation observables in relativistic heavy-ion collisions.
First, we discuss the effects of 
the thermal blurring, i.e. the blurring effect arising from 
the experimental measurement of fluctuations in momentum space
as a proxy of the thermal fluctuations defined in coordinate space, 
on higher order cumulants.
Second, we discuss the effect of imperfect efficiency of detectors 
on the measurement of higher order cumulants. 
We derive effective formulas which can carry out the 
correction of this effect for higher order cumulants
based on the binomial model.
\end{abstract}

  
\section{Introduction}

Fluctuations of conserved charges are important observables
in relativistic heavy-ion collisions because they are 
believed to be sensitive to early thermodynamics of 
the hot medium created by the collisions \cite{Asakawa:2015ybt}.
Active theoretical and experimental studies on the fluctuation 
observables as well as numerical simulations on the lattice and 
their comparison have been carried out.
In particular, non-Gaussianity of fluctuations characterized by 
higher-order cumulants acquires much attention recently.
For example, the sign of higher-order cumulants is believed to 
be sensitive to the location of a medium in the QCD phase diagram 
\cite{Asakawa:2009aj,Stephanov:2011pb}.

In theoretical studies and lattice simulations on fluctuation observables, 
they are usually studied based on statistical mechanics. 
The fluctuations discussed in this formalism are called thermal fluctuations.
On the other hand, the experimental analyses measure the event-by-event
fluctuations of particle number observed by detectors.
Although these fluctuations are sometimes compared directly in 
the literature, the experimentally-observed fluctuations are not the
same as the thermal fluctuations in various aspects.
In this talk, we focus on two such differences
\cite{Ohnishi:2016bdf,Kitazawa:2016awu}, and clarify the reason why 
they are different and in which case they can be compared with each other.
We also discuss practical ways to compare these different 
fluctuations.

\section{Thermal blurring}

\begin{figure}
\begin{center}
\includegraphics[width=.55\textwidth]{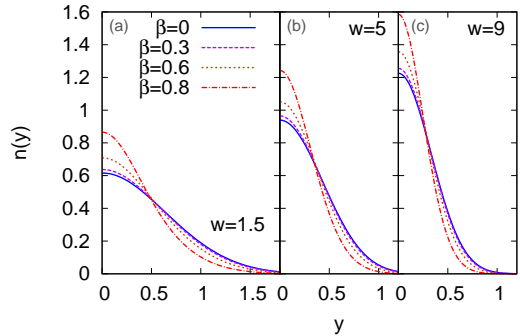}
\caption{
Particle number distribution per unit rapidity $n(y)$
for several values of $w=m/T$ and the transverse velocity $\beta$
\cite{Ohnishi:2016bdf}.
}
\label{fig:y-dist}
\end{center}
\end{figure}

We first focus on the difference in phase spaces in which thermal 
fluctuations and the fluctuation observed in experiments are defined
\cite{Ohnishi:2016bdf}.
First, thermal fluctuations calculated in statistical mechanics are
those in a spatial volume with momentum space integrated out.
On the other hand, the experimental detectors can only measure 
the momentum of particles entering there. 
Therefore, the event-by-event fluctuations measured in experiments 
are inevitably those in a momentum phase space.
Therefore, the phase spaces defining fluctuations are completely 
different between theoretical and experimental settings.

Nevertheless, for sufficiently high energy collisions
the latter can be regarded as a proxy of the former 
as recognized in earlier studies \cite{Asakawa:2000wh,Jeon:2000wg}.
This approximate correspondence can be obtained by assuming the 
Bjorken space-time evolution.
In this space-time picture, the system has boost invariance.
Accordingly, (momentum-space) rapidity and coordinate-space rapidity,
\begin{eqnarray}
y=\frac12 \ln \frac{ E-p_z }{ E+p_z }, \quad
Y=\frac12 \ln \frac{ t-z }{ t+z }, 
\end{eqnarray}
of a fluid element are equivalent in this picture, 
where $z$ represents the longitudinal coordinate.
If all particles were stopping in each fluid element,
therefore, the experimental measurement in a rapidity window
$\Delta y$ can be regarded as the one in the coordinate-space
rapidity window with $\Delta Y= \Delta y$.

However, particles have thermal motion, and thus they have
nonzero velocity in the rest frame of a fluid element.
The correspondence between $y$ and $Y$ thus is only 
an approximate one for individual particles.
Due to this difference, the experimental measurement of a particle number 
in $\Delta y$ receives a blurring effect when it is interpreted
as the coordinate-space one in $\Delta Y$.
We call this effect as thermal blurring.
In Fig.~\ref{fig:y-dist}, we show the thermal distribution 
of a free particle with mass $m$ in rapidity space in a medium
with temperature $T$ and radial velocity $\beta$.
The result is plotted for various values of $w=m/T$.
The blastwave model for the central collisions at LHC and top-RHIC 
energies suggests $T\simeq100$ MeV and $\beta=0.6-0.7$ at 
kinetic freezeout; thus pions have $w\simeq1.5$ 
while nucleons have $w\simeq9$ at kinetic freezeout.
From Fig.~\ref{fig:y-dist}, one finds that for pions the 
thermal width in $y$ space is as large as $1.0$, which is 
comparable with the maximum coverage of the STAR detector $\Delta y=1.0$.
This result indicates that the measurement of net-electric charge 
fluctuation at STAR is strongly affected by the thermal blurring
effect.
When the event-by-event fluctuation at $\Delta y=1.0$ is compared
with the thermal one, therefore, the thermal blurring effect has to 
be considered seriously.

In Ref.~\cite{Ohnishi:2016bdf}, it is discussed that the 
thermal blurring effect gives rise to a characteristic 
dependence of higher-order cumulants as a function of $\Delta y$.
This result suggests that the experimental result of the cumulants 
as a function of $\Delta y$ can be used to understand and remove 
the thermal blurring effect.
It is also discussed that the thermal blurring can be regarded as
a part of diffusion before kinetic freezeout \cite{Kitazawa:2015ira}.
It is an interesting subject to investigate the $\Delta y$ dependences
of higher-order cumulants experimentally, and extract the values of 
cumulants before the blurring and diffusion effects.
See Ref.~\cite{Kitazawa:2015ira} for more detailed discussion.

\section{Efficiency correction}

Next, we consider the effect of the imperfect efficiency of detectors
on fluctuation observables \cite{Kitazawa:2016awu}.
All detector can measure particles entering there only with a 
probability less than unity, which is called the efficiency.
Because of the finite efficiency, the event-by-event distribution
of a particle number observed experimentally is different from 
the original one without the efficiency loss.
The cumulants characterizing the distribution function, therefore,
are also modified by the efficiency loss.
This difference has to be corrected in experimental analyses
to remove the artificial effect due to the efficiency loss.

It is known that this procedure for efficiency correction 
can be carried out when the efficiencies for individual particle 
are independent \cite{Kitazawa:2012at}.
In this case, the event-by-event probability distribution functions 
of observed and original particle numbers are related 
with each other in a simple form with binomial distribution functions
\cite{Kitazawa:2012at}.
Using this relation, the cumulants of original particle number 
without efficiency loss can be represented by those of observed 
particle numbers with efficiency loss.
These relations are first obtained for net-particle number 
with two different efficiencies for particle and anti-particle 
\cite{Kitazawa:2012at}.
The relation is then extended to the case that the particle 
number is given by a sum of the numbers of different species of 
particles which are observed with different efficiencies
\cite{Luo:2014rea,Bzdak:2013pha}.

The procedure of the efficiency correction presented in 
Refs.~\cite{Luo:2014rea,Bzdak:2013pha} is given in terms of
factorial moments.
In this method, however, the number of factorial moments grows
as the number of different particle species $M$ increases.
Accordingly, the numerical cost required for 
the efficiency correction becomes large rapidly with 
increasing $M$;
for $n$-th order cumulant, the cost increases $\sim M^n$.
In order to carry out the efficiency correction with a reasonable
computational time, therefore, the number of $M$ is limited to 
a small number.

In Ref.~\cite{Kitazawa:2016awu}, different formulas which lead
to the same result as those in Refs.~\cite{Luo:2014rea,Bzdak:2013pha} 
are obtained.
In this formula, the efficiency correction of a charge which is given by 
\begin{eqnarray}
Q = \sum_{i=1}^M a_i N_i ,
\label{eq:Q}
\end{eqnarray}
is considered, where $N_i$ is a particle number with a species 
of particle labeled by $i$, and $a_i$ is a numerical number.
Because of the efficiency loss, the number $n_i$ observed 
by the detector can be different from, and typically smaller than,
$N_i$.
The cumulants of $Q$ up to fourth order are given by
\begin{eqnarray}
\langle Q \rangle_{\rm c} 
&=& \llangle q_{(1)} \rrangle_{\rm c} ,
\label{eq:Q1}
\\
\langle Q^2 \rangle_{\rm c} 
&=& \llangle q_{(1)}^2 \rrangle_{\rm c} - \llangle q_{(2)} \rrangle_{\rm c} ,
\label{eq:Q2}
\\
\langle Q^3 \rangle_{\rm c} 
&=& \llangle q_{(1)}^3 \rrangle_{\rm c} - 3 \llangle q_{(2)} q_{(1)} \rrangle_{\rm c} 
+ \llangle 3 q_{(2,1|2)} - q_{(3)} \rrangle_{\rm c} ,
\label{eq:Q3}
\\
\langle Q^4 \rangle_{\rm c} 
&=& \llangle q_{(1)}^4 \rrangle_{\rm c} - 6 \llangle q_{(2)} q_{(1)}^2 \rrangle_{\rm c} 
+ 12 \llangle q_{(2,1|2)} q_{(1)} \rrangle_{\rm c} 
+ 3 \llangle q_{(2)}^2 \rrangle_{\rm c} 
\nonumber \\
&&
-4 \llangle q_{(3)} q_{(1)} \rrangle_{\rm c} 
+ \llangle -15 q_{(2,2|2)} + 6 q_{(2,1,1|3)} + 4 q_{(3,1|2)} 
- q_{(4)} \rrangle_{\rm c} ,
\label{eq:Q4}
\end{eqnarray}
where the cumulants $\langle\cdot\rangle_{\rm c}$ and 
$\llangle\cdot\rrangle_{\rm c}$ are taken 
for original distribution function for $N_i$ and observed one 
for $n_i$, respectively.
$q_{(\cdots)}$ are linear combinations of $n_i$
defined by
\begin{eqnarray}
q_{(s)} 
= \sum_{i=1}^M c_{(s)}^{(i)} n_i,
\qquad
q_{(s_1,\cdots,s_j|t_1,\cdots,t_k)} 
= \sum_{i=1}^M c_{(s_1,\cdots,s_j|t_1,\cdots,t_k)}^{(i)} n_i.
\label{eq:q(sstt)}
\end{eqnarray}
The coefficients $c_{(\cdots)}^{(i)}$ are numerical numbers 
which depend on $a_i$ and $\epsilon_i$ defined by
\begin{eqnarray}
c_{(s)}^{(i)} = \tilde{a}_i^s \tilde\xi_s^{(i)} ,
\qquad
c_{(s_1,\cdots,s_j|t_1,\cdots,t_k)}^{(i)}
= \tilde{a}_i^{s_1+\cdots+s_j} \tilde\xi_{s_1}^{(i)} \cdots \tilde\xi_{s_j}^{(i)} 
\tilde\xi_{t_1}^{(i)} \cdots \tilde\xi_{t_k}^{(i)} ,
\end{eqnarray}
with 
$\tilde{a}_i = a_i/\epsilon_i$,
$\tilde{\xi}_m^{(i)}=\xi_m^{(i)}/\epsilon_i$, and 
$\xi_m^{(i)}=\xi_m(\epsilon_i)$ being coefficients of 
the binomial cumulants with probability $\epsilon_i$.

Equations~(\ref{eq:Q1}) -- (\ref{eq:Q4}) represent the cumulants
of the original particle number $Q$ by the 
mixed cumulants of observed particle numbers.
Since the right-hand sides in these equations are experimental 
observables, these equations enable the efficiency correction.
The formulas Eqs.~(\ref{eq:Q1}) -- (\ref{eq:Q4}) consist of 
a fixed number of mixed cumulants.
Because of this property, the numerical cost for the 
efficiency correction is proportional to $M$ for all orders 
of cumulants.
This is contrasted to the numerical cost of the method in 
Refs.~\cite{Luo:2014rea,Bzdak:2013pha} which is proportional 
to $M^n$.
The new formula can drastically reduce the numerical cost 
especially when $M$ and $n$ are large.
Because of this advantage, Eqs.~(\ref{eq:Q1}) -- (\ref{eq:Q4}) 
will enable us to carry out the efficiency corrections
with momentum dependent efficiency.
It is an important subject to perform such efficiency corrections
in experimental analyses.

\end{document}